# Programmers Aren't Obsolete Yet: A Syllabus for Teaching CS Students to Responsibly Use Large Language Models for Code Generation

Bruno Pereira Cipriano[1][a] and Lúcio Studer Ferreira[1][b]
[1]*Lusófona University, Portugal*
{*bcipriano, lucio.studer*}@*ulusofona.pt*

Abstract: Large Language Models (LLMs) have emerged as powerful tools for automating code generation, offering immense potential to enhance programmer productivity. However, their non-deterministic nature and reliance on user input necessitate a robust understanding of programming fundamentals to ensure their responsible and effective use. In this paper, we argue that foundational computing skills remain crucial in the age of LLMs. We propose a syllabus focused on equipping computer science students to responsibly embrace LLMs as performance enhancement tools. This work contributes to the discussion on the why, when, and how of integrating LLMs into computing education, aiming to better prepare programmers to leverage these tools without compromising foundational software development principles.

## 1 INTRODUCTION

Large Language Models (LLMs) have been shown to have the capacity to generate computer code from natural language specifications (Xu et al., 2022; Destefanis et al., 2023). Currently, there are multiple available LLM-based tools that display that behavior. Two examples of such tools are OpenAI's ChatGPT [1] and Google's Gemini [2].

While these tools have the potential to improve productivity in software development and other computer science and engineering related fields, they also pose obvious threats, since students now have access to tools that can generate code to solve a variety of programming assignments. If nothing is done, universities are at risk of producing low-quality graduates who do not possess the skills required to succeed in the job market.

This has lead to multiple discussions in the Computer Science Education (CSE) community on how to deal with these tools. Some educators decide to fight them, while others prefer to embrace them into their workflows and classes (Lau and Guo, 2023).

Several trends of work have explored ways to mitigate the impact of these models and reduce student's overreliance on them. One example is the use of diagram-based exercises, which force students to interpret a visual description of a problem, thus reducing the students' ability to obtain solutions by mere "copy-and-prompting" (Denny et al., 2023a; Cipriano and Alves, 2024b).

On the other hand, LLMs are becoming part of industry practice (Barke et al., 2022; DeBellis et al., 2024), and interacting with these tools in a professional capacity will likely become more prevalent as time goes by, due to the expected productivity gains.

As such, in this position paper, we argue for the creation of LLM-code-generation training within computer science and computer engineering courses. Furthermore, we propose a set of theoretical topics and practical activities as a first approach to a syllabus for such a course.

This paper makes the following contributions:

- Arguments for adopting a new course, "Responsible Software Development using Large Language Models", tailored for Computer Science and Engineering programs;

- Arguments for the importance of learning the classical computer science skills before using LLMs for code generation;

- Presents theoretical topics and practical activities to equip students with the knowledge and skills to responsibly use LLMs as software development support tools.

---

[a]  https://orcid.org/0000-0002-2017-7511
[b]  https://orcid.org/0000-0002-0235-2531
[1] https://chat.openai.com/
[2] https://gemini.google.com/app

## 2 FUNDAMENTAL CONCEPTS AND RELATED WORK

### 2.1 Fundamental Concepts

This section introduces key concepts necessary for understanding the remainder of the paper.

- Prompt: A natural language description of a task to be performed by the LLM.
- Prompt Engineering: The process of crafting prompts to enhance LLM performance, such as using role-playing techniques (e.g., "From now on, you are a Math teacher. Please answer the following Math question" (Kong et al., 2023)).
- Hallucinations: The phenomenon where LLMs generate incorrect, misleading, or fabricated information (e.g., nonexistent book titles or authors (Daun and Brings, 2023)).
- Copy-and-prompting: Directly inputting assignment instructions into an LLM (Cipriano et al., 2024), often bypassing meaningful effort or critical engagement.

### 2.2 Students' Interaction with LLMs

In (Babe et al., 2023), researchers asked 80 students with a single semester of programming background to write prompts for 48 problems. They found that the students' prompts can be ambiguous to LLMs, leading to the generation of multiple semantically different functions. This study provides an example prompt which, while being clear for humans to interpret, resulted in 7 semantically different functions being generated by the LLM StarCoderBase.

The authors of (Cipriano and Alves, 2024a) ran a student survey in order to evaluate 1st-year CS students' opinions on the academic and professional usage of LLMs. They reported that a combined 76.9% of students indicated that it would be 'helpful' (25%) or 'very helpful' (51.9%) if teachers would teach them how to use GPT more effectively. Furthermore, 23.1% of the students reported needing 'many prompts' in order to get useful results from GPT and 1.9% reported that they 'usually can't get useful results'. Finally, a combined 88.5% of the students indicated that having GPT-oriented exercises in their courses would help them in their professional futures.

In (Smith et al., 2024), researchers surveyed a total of 133 CS students from multiple levels (166 undergraduate students, 12 masters' students and 5 PhD students). Among other findings, they reported that 53 students (45.69%) indicated a desire for being taught "professional usage of GenAI", and that 30 students (19.87%) indicated a desire for the "integration of GenAI in the curriculum". Many students (41, or 24.70%) justified their position by indicating that they anticipate the professional usage of these tools in their career. Students also voiced some concerns caused by these tools, such as worries about misinformation, unethical uses, intellectual property violations privacy breaches, unfair advantages or equitable access issues and even job replacement.

In (Alves and Cipriano, 2024), educators asked students to interact with ChatGPT during the scope of a data-structures and algorithms (DSA) project, and analyzed the resulting interaction logs. The DSA project had a specific requirement which should be implemented using ChatGPT and for which the instructors provided a template which required students to obtain two alternative solutions from the LLM, and provide a written comparison between them, with the goal of fostering students' critical thinking. Among other findings, they reported that 1) a significant portion of students (38.9%) failed to ask for the second solution, and, 2) students prompts tended to be unsophisticated, lacking context, examples or even the expected function signature. These authors also present some case studies selected from their analysis, with one of them showing a student who included the word 'database' in the respective prompt, leading GPT to generate SQL code which was outside the scope of the DSA project. This student is also reported as not being able to integrate GPT's solution in their own project.

In conclusion, previous research shows that 1) not all students are able to take advantage of these tools for code generation, 2) students are interested in learning how to better use these tools, and 3) students believe that LLMs will be an important part of their professional future.

### 2.3 Integration of LLMs in Programming Education

In (Denny et al., 2023a; Denny et al., 2024), educators describe a new pedagogical approach which tries to reduce LLM-overreliance by presenting introductory programming students with diagram-based exercises instead of the classical text-based ones. Furthermore, to solve those exercises, students are expected to craft LLM-prompts which are then submitted to an LLM to obtain working code. Finally, the LLM-generated code is tested by teacher-developed unit tests. Similarly, the authors of (Cipriano et al., 2024) have proposed a diagram-and-video-based approach for prescribing Object-Oriented Programming (OOP) exercises. Besides countering LLM-overreliance, these approaches force students to interact to LLMs in more authentic ways than mere "copy-and-prompting" of the assignments' instructions.

The authors of (Reeves et al., 2024) suggest that natural language programming is the expected evolution of programming, reflecting back to a paper from 1966 by Mark Halpern which mentioned that *"natural programming language is one that can be written freely, not just read freely"*. These authors argue that LLMs have earned their place in programming education, with the main question being *when to teach students how to use them*. Finally, they advocate for the definition of a 'precise programming vocabulary' to facilitate students' communication with LLM-based code generation tools in a non-ambiguous way, similarly to what happens in fields such as Mathematics, and for a higher focus on the development of debugging skills, due to the likelihood that these non-deterministic tools generate incorrect, buggy and/or insecure code (Cipriano and Alves, 2024b; Asare et al., 2023).

In (Vadaparty et al., 2024), educators describe an experiment in which they allowed students to use GitHub Copilot in a CS1 course, starting from the first class. These authors adapted their previous course, introducing new topics, such as problem decomposition and prompt engineering, and giving higher importance to previously existing topics such as testing and debugging. Although these educators report some positive aspects, such as students being able to implement more complex software than before, they also reported some difficulties. One of those difficulties was related with the introduction of LLMs in the first week of the course. This difficulty led those educators to propose delaying the introduction of the LLM until the students are able to implement "small programs" on their own. Another issue was related with students' confusion in terms of what they should be able to do with and without resorting to the LLM. To overcome this issue, the authors recommend that each assignment is tagged in order for students to clearly differentiate what should be with and without the LLM.

More recently, the authors of (Keuning et al., 2024) suggested an approach where CS1 (i.e., the course where computer programming is traditionally first introduced to CS students) is taught in what can be seen as an 'inverse mode'. This approach reimagines the teaching sequence by starting with (1) brainstorming, requirements analysis, formulation of user stories and breaking functionality into smaller components, (2) progressing to User Interface design and software design, followed by (3) to code snippet generation, code improvement, code composition (i.e., connecting components) and debugging, (4) testing, and finally, (5) deploying the software. These authors also identified several critical questions that must be addressed before implementing such as curriculum change. Among these are: *What prerequisite skills do students need [before entering such a course]?*, and *In the context of code-writing, when should students start using GenAI tools?*

In conclusion, this research avenue is still in its early stages, but some common topics, such as the need for better training in software validation (e.g., debugging, testing), appear to be consensual due to the non-deterministic nature of these tools.

## 3 FOUNDATIONAL COMPUTING SKILLS FOR EFFECTIVE AND RESPONSIBLE LLM USE

We believe that certain foundational skills must be acquired before students can effectively (and responsibly) use LLMs for code generation. A useful analogy can be drawn between elementary math skills (addition, subtraction, multiplication, and division) and calculators. The global education community did not abandon teaching basic math when calculators became widely available. Even in more advanced topics, such as graphing functions, calculators are used to complement learning, but students are still expected to understand how to plot a function manually. This foundation ensures they can critically evaluate the tool's output. While calculators are deterministic and generally reliable, they can still produce incorrect results in rare cases, such as when bugs exist in their programming. In contrast, LLMs are inherently non-deterministic and can frequently generate incorrect or misleading outputs (i.e., hallucinations). Given this, it is even more critical for students to grasp the basic principles behind the code that LLMs produce. Just as students must understand the operations underpinning calculators, they should also understand the logic and structure behind the outputs of LLMs to use these tools effectively and responsibly.

The need for foundational skills can also be understood through the lens of Bloom's Taxonomy (Bloom et al., 1964), which categorizes cognitive processes into a hierarchy. Foundational skills like understanding basic syntax or logical structures (i.e., variables, data-types, if/then/else, loops, functions, etc.) correspond to the lower-order cognitive levels of **Remembering** and **Understanding**. These levels form the essential groundwork for students to progress toward higher-order thinking, such as **Applying** their knowledge in novel situations, **Analyzing** the correctness of

LLM outputs, and **Evaluating** or refining these outputs for practical use. Ultimately, these skills enable students to reach the pinnacle of the taxonomy: **Creating**, where they can confidently design their own solutions and innovations, using LLMs to improve their development efficiency. Skipping the foundational stages risks leaving students ill-prepared to engage in these higher-level processes, especially when using a tool as unpredictable as an LLM.

Some authors suggest that LLMs represent a new step in the abstraction chain, akin to how higher-level languages reduced the need to master concepts like memory management and data types (Reeves et al., 2024). However, current LLMs function more as "translators" which can convert natural language instructions into existing programming languages (e.g., Java and Python) rather than introducing new programming paradigms or higher-level constructs. This distinction underscores the continued importance of mastering foundational programming skills, as LLMs rely on the programmer's ability to specify precise instructions and critically evaluate the generated outputs. Without a solid grasp of syntax, logic, and basic programming constructs, users risk becoming overly reliant on these tools while lacking the knowledge to verify or adapt their results effectively.

### 3.1 Pre-requisite knowledge

In our opinion, before entering a course focused on using LLMs for code generation, students should have had at least the following academic experience:

- One semester of an introductory course covering the basics of variables, data-types, if/then/else, loops, functions and problem decomposition, such as typically taught in CS1 courses;
- Knowledge of data-structures such as arrays, lists, queues, and stacks, which are usually covered in CS1 and/or CS2 courses;
- Elementary knowledge of searching – e.g., linear and binary search –, and sorting algorithms – e.g., bubble sort, selection sort, merge sort, and quicksort – which are typically presented in CS2 courses;
- Notions of algorithmic efficiency and/or complexity, such as Big-O notation, which are typically covered in CS2 or Data-Structures and Algorithms courses;
- Knowledge of modeling and design, including domain analysis, modularity, abstraction, and relationships between entities. These skills are typically developed in courses such as OOP (e.g., inheritance and composition) and Databases (e.g., entity-relationship model).

Why are these pre-requisites relevant?

The **introductory course** will teach students the basic programming constructs (logic, variables, operators, loops, functions), problem decomposition and problem solving. It will also give students the basic ability to read and understand code. This elementary knowledge will allow students to interpret what the generated code is doing and to reason about its capabilities and limitations. A student with this basic knowledge will be able to identify bugs and improvements such as: *"This code will result in an integer overflow, thus causing the calculation to be incorrect. I will instruct the LLM to support the case where the sum of the numbers in the array does not fit a 32-bit integer."*

Knowledge of elementary **data-structures**, **searching and sorting** techniques, along with notions of **algorithmic efficiency and/or complexity**, will enable students to identify and/or handle situations where the generated code is not efficient enough to solve the problem. A student informed about these topics can analyse the generated code and apply thought processes such as: *"The generated function includes a sorting algorithm with an average-case complexity of $O(N^2)$ which is incompatible with the amount of data that we need to process; I'll clarify the problem's constraints and ask the LLM to use a more efficient algorithm."*

Having knowledge of **modeling and design principles** enables students to critically assess whether LLM-generated artifacts (e.g., code or database schemas) accurately represent the problem domain. This includes identifying missing or misrepresented entities, relationships, or constraints, as well as evaluating maintainability and adaptability to future changes. A student equipped with this knowledge can analyse solutions and apply thought processes such as: *"A key domain concept was not represented as a class. I will ask the LLM to take give more importance to that concept."*, and *"There is some repetition between these two concepts, I will instruct the LLM accordingly."*

## 4 COURSE PROPOSAL

We propose the development of a new course, "Responsible Software Development using Large Language Models", designed to teach students how to responsibly approach and use LLMs for code generation.

The primary goals of the course are:

- Equip students with the required knowledge to safely leverage LLMs as productivity-enhancing software development tools;
- Promote ethical use by fostering awareness of biases inherent in LLM training data and their potential to produce prejudicial system outcomes.

### 4.1 Syllabus

The proposed course combines theoretical and practical classes, structured into four main topics: (1) Introduction to LLMs, (2) LLMs for Software Developers, (3) Prompt Engineering, and (4) Validation Techniques.

The following four subsections detail each of these topics, while a fifth subsection is dedicated to the practical activities that complement the theoretical content.

**Introduction to LLMs**

This topic introduces students to the evolution of Natural Language Processing, the inner workings of LLMs (e.g., their probabilistic nature, tokenization, etc.), their sensitivity to training data, and other related concepts.

Students will also be exposed to techniques which can be used to improve the models' performances, such as prompt-engineering, 'fine-tuning'[3] and Retrieval-Augmented Generation (RAG)[4].

Furthermore, students will be made aware of relevant ethical considerations, such biases that the tool's might have and potential intellectual property and copyright issues that can arrive from using these tools, both as data consumers (i.e., LLM-users) as well as data producers (e.g., when sharing their own code with a non-local LLM).

Finally, students will be encouraged to apply critical thinking when handling these tools (similarly to the proposal of (Naumova, 2023) for training health professionals). They will be taught the importance of validating and double-checking their results and to check alternative information sources.

**LLMs for Software Developers**

This topic will delve into subjects more closely related with programming and software development. Students will be taught the main differences between general purpose / dialog-oriented models (e.g., ChatGPT, Gemini, Claude, and Llama) and Code Models (e.g., Code Llama and StarCoderBase), in terms of advantages, disadvantages, relative strengths, cost, access-types (i.e., executed locally, web-based, API-based) as well as other relevant criteria. They will also be introduced to tools which integrate LLMs in Integrated Development Environments (IDEs), such as GitHub Copilot and GitHub Copilot Chat.

**Prompt Engineering**

Students will be presented with prompt engineering techniques of demonstrated value such as Chain-of-Thought (Wei et al., 2022), Role Playing (Kong et al., 2023), and Self-Consistency (Wang et al., 2023).

**Validation Techniques**

Students will learn techniques for validating LLM-generated code outputs, such as debugging, testing, logging, and code review. They will also be introduced to a taxonomy of automated testing types (e.g., unit, integration, end-to-end) along with their use cases and scenarios. Additionally, students will be cautioned against using LLMs to generate unit tests for code also generated by LLMs, as this approach can lead to biased and potentially incorrect tests—a concern highlighted by other researchers (Korpimies et al., 2024; Espinha Gasiba et al., 2024).

**Practical Activities**

The course combines hands-on activities to develop two skill types: solution validation and LLM usage. Initially, validation-focused activities teach students to assess and improve LLM-generated solutions without relying on LLMs. These activities are detailed in Table 1. Once students gain experience with validation, they progress to activities aimed at building LLM-usage skills, as outlined in Table 2.

### 4.2 Evaluating Student Performance

The theoretical component and the solution-validation skills (e.g., unit testing, debugging) can be evaluated using written exams or using exercises which measure the impact of the techniques (e.g., measure the test coverage percentage).

Evaluating students' interactions with LLMs is harder, due to these tools' non-deterministic nature. As such, this evaluation should be mostly focused on the process, considering also the results. For each interaction evaluation, students should deliver a report which documents their thought process and presents a log of the used prompts. This report will then be qualitatively evaluated by an instructor, who will assess whether the student has applied the expected validation techniques. For example, the evaluation will consider whether unit tests were written to validate the LLM's output and whether a critique of the LLM's solutions was presented. This process-oriented evaluation aligns with recommendations from other researchers (Prather et al., 2023; Feng et al., 2024).

---

[3] Fine-tuning: process of retraining an LLM on task- or domain-specific data to improve its performance on specialized tasks.

[4] RAG: the process of enhancing an LLM's output by automatically retrieving relevant documents from an external database and incorporating them into the prompt to provide additional context.

Table 1: Validation techniques: activities, descriptions, and learning outcomes.

| Activity | Description | Learning Outcome |
|---|---|---|
| Code reading / explaining | Engage in code reading exercises of increasing complexity, starting from single-function to multiple-class exercises. They will produce reports detailing the temporal evolution of variables and program outputs. | Practice reading code produced by others, a skill valued by the industry and with increased importance when dealing with AI-generated code (Denny et al., 2023b). |
| Debugging | Practice the use of debugging tools, including stepping through code, setting breakpoints, and inspecting variables to find facts about code, ranging from single-function to multi-class scenarios. | Acquire skills to debug efficiently, trace variable changes, and apply debugging concepts to complex scenarios involving multiple components. |
| Test Driven Development (TDD) | Define and implement a test battery for a function/class that implements a functional goal. Afterwards, ask the LLM to implement the function/class and validate the generated code using the test battery. The tests must be implemented without using an LLM. | Learn how to use TDD to guide, validate and increase the quality of LLM-generated solutions. |
| Integration Testing | Define and implement integration tests for a program composed of multiple classes. The tests must be implemented without using an LLM. | Learn to conceptualize and implement tests involving multiple code components, enabling the testing of integration and composition between of various LLM-generated code portions. |
| Logging | Applying logging to an existing code base. | Learn how to use a real-world logging package, such Python's `logging` package. |
| Code review | Students will pair up to discuss the pros and cons of code samples, simulating a real-world code review. | Develop a mindset of evaluating and criticizing code produced by others. |

## 5 DISCUSSION

Reflecting on the work of (Keuning et al., 2024), which posed critical questions such as "*What are the pre-requisites?*" and "*When should LLMs be introduced for code generation?*", we presented our perspective in Section 3, "Foundational Computing Skills for Effective LLM Use". We outlined a set of foundational courses and knowledge that we consider essential before students begin using LLMs to produce working software. Specifically, we argue that students must first master fundamental computational skills and demonstrate the ability to solve medium-complexity programming problems (e.g., involving 2-3 classes) independently.

While we define these skills and courses as prerequisites for an LLM-focused code generation course, we also recognize the potential benefits of exposing students to LLMs during these foundational courses. However, it is crucial that such exposure does not compromise the learning of core programming basics. This can be ensured through measures such as in-person, proctored evaluations, where students individually modify their assignment code without relying on LLMs.

We anticipate some difficulties with the development of this course, due to LLM's non-deterministic nature as well as their continuous development (i.e., their behaviours are changing over time (Chen et al., 2023)). First, LLMs' output for the same task or exercise might change between sessions or updates, po-tentially causing issues in live demos. One option to work around this problem is to follow the recommendations of (Vadaparty et al., 2024) and use static slides with previously obtained LLM-responses which highlight significant issues. Although that technique solves the problem of presenting and discussing theoretical information, it is insufficient for more practical exercises. Designing practical exercises poses additional challenges, as it is difficult to create code-generation tasks for which the LLM consistently produces either strong or flawed outputs. A possible approach is to identify topics where LLMs typically struggle (as done, for example, in (Cipriano and Alves, 2024b)) and build exercises around those topics, while also ensuring that they are relevant for the learning goals.

Table 2: LLM usage skills: activities, descriptions, and learning outcomes.

| Activity | Description | Learning Outcome |
| --- | --- | --- |
| Brainstorming | Use dialog-oriented models (e.g., Copilot Chat, ChatGPT, Gemini) to generate and refine ideas, such as clarifying requirements, creating user stories, and defining acceptance criteria. | Develop skills to effectively interact with dialog-oriented LLMs for brainstorming, refining ideas, and defining clear, actionable goals through iterative dialogue. |
| Find-the-Bug Exercises | Starting with a buggy code base, use an LLM to find and fix the problems. | Learn how to use LLMs to debug and improve code quality. |
| Critical Comparison of LLM-generated code | Generate multiple solutions, analyze them, identify issues, select one, justify the choice, and suggest improvements. | Develop critical thinking skills by comparing, evaluating, and improving LLM-generated code solutions. |
| Guiding LLMs towards better solutions | Use examples (i.e., test cases) and restrictions (e.g., code style rules, preferred or forbidden libraries/keywords, time/space complexity constraints, etc) to guide LLMs in generating improved solutions. | Develop the ability to systematically interact with LLMs in order to obtain solutions with varying properties. |
| Group Discussions | Engage in class discussions focused on prompts for achieving specific programming goals. | Gain insights from peers' ideas and experiences with prompt creation for code generation. |
| LLM-based existing code exploration | Use an LLM to explore and document an existing but unknown code base, which can be teacher-supplied or developed by other students in a prior course. Following this exploration, the students can be asked to augment the project's functionality using LLMs. | Gain experience using an LLM to understand a sizeable, unfamiliar code base, simulating real-world scenarios often encountered in the industry. |

Another challenge is the evaluation of the students' performance and acquired knowledge. For the theoretical component and for the validation techniques this is straightforward. However, that is not true for the LLM-interaction components. We have decided to qualitatively evaluate that part of the course, focusing on processes over product, at least for now. Still, we recognize that this approach does not scale well and may require adjustments in the future. Other, more quantitative possibilities exist, such as measuring the number of prompts needed to reach a certain goal, or validating the generated code using teacher-defined unit tests, similarly to what is proposed by (Denny et al., 2024). Although these alternatives are interesting, they might lead to unfair results, since students results will be somewhat dependent on the tools non-deterministic output.

The ideal placement of this course within the curricula remains uncertain. Based on the typical sequencing of prerequisite courses, we believe it should be offered after the third semester. However, it may be more appropriate to schedule it in the third year, as students at that stage tend to have a more mature understanding of software development principles. This increased maturity may make them more receptive to learning the proposed validation techniques.

We acknowledge that this course may become obsolete in a few years, either due to the continuous evolution of LLM capabilities or their better integration into prerequisite courses. However, this does not diminish its relevance and necessity during this transitional period since it is crucial to equip students with the skills to critically engage with these tools rather than passively rely on them.

Finally, we believe that the proposed course may serve as a laboratory for experimentation, potentially generating insights and approaches that could be adapted for use in other courses, including prerequisite ones. In fact, the very process of refining effective activities and exercises for integrating LLMs without compromising foundational programming skills may contribute to the eventual obsolescence of the course itself, as these best practices become embedded in earlier stages of the curriculum.

# 6 CONCLUSIONS

We believe that it is of the upmost importance to integrate LLM-related training in the curricula of computer science and engineering degrees. These tools are already being used in the industry (DeBellis et al., 2024), which means that companies will expect graduates to have authentic experiences with the use of these tools to produce working software. Although some courses worldwide are already doing integrations of these tools (as seen, for example, in (Korpimies et al., 2024)), we believe that the topic's importance and depth justifies a new curricular unit.

Although further research is still needed to develop the optimal pedagogical approaches and techniques for integrating LLMs in computer science education, the sooner the CSE community starts to integrate these tools, the sooner we will be able to develop those approaches and techniques.

With this work we hope to contribute to the general discussion of *why*, *how*, and *when* students should be exposed to LLMs as software development support tools. We have plans to begin implementing a pilot version of the described course in the second semester of the 2024/2025 school year. We will share any relevant findings with the CSE community.

# ACKNOWLEDGEMENTS

This research has received funding from the European Union's DIGITAL-2021-SKILLS-01 Programme under grant agreement no. 101083594.